\documentclass{aa}
\usepackage{graphicx}
\usepackage{txfonts}
\newcommand{\etal}{et\,al.\ }
\newcommand{\logg}{\mbox{$\log g$}}
\newcommand{\Teff}{\mbox{$T_\mathrm{eff}$}}
\newcommand{\pgstar}{\object{HE1429$-$1209}}
\newcommand{\dostar}{\object{HE1314$+$0018}}

\newcommand{\rxj}{\object{RX\,J2117.1$+$3412}}

\newcommand{\ngc}{\object{NGC\,246}}

\newcommand{\keins}{\object{K1$-$16}}

\begin{document}
   \title
   {Identification of a DO white dwarf and a PG1159 star in the ESO SN~Ia progenitor survey (SPY)\thanks
{Based on observations at the Paranal Observatory of the European Southern
   Observatory for programs No.\ 165.H-0588(A) and 167.D-0407(A).}
   }
 
   \author{K. Werner$^1$, T. Rauch$^{1,2}$, R. Napiwotzki$^3$,
   N. Christlieb$^4$, D. Reimers$^4$, and C.A. Karl$^2$}
   \offprints{K\@. Werner}
   \mail{werner@astro.uni-tuebingen.de}
 
   \institute
    {
     Institut f\"ur Astronomie und Astrophysik, Universit\"at T\"ubingen, Sand 1, D-72076 T\"ubingen, Germany
\and
Dr.-Remeis-Sternwarte, Universit\"at Erlangen-N\"urnberg, Sternwartstra\ss e 7, D-96049 Bamberg, Germany
\and
Department of Physics and Astronomy, University of Leicester, University
    Road,  Leicester, LE1 7RH, UK
\and
Hamburger Sternwarte, Universit\"at Hamburg, Gojenbergsweg 112, D-21029 Hamburg, Germany
}
    \date{Received xxx / Accepted xxx}
   \authorrunning{K. Werner et al.}
   \titlerunning{Identification of a DO white dwarf and a PG1159 star in the SPY}
   \abstract{We present high-resolution VLT spectra of a new helium-rich DO
   white dwarf (\dostar) and a new PG1159 star (\pgstar), which we identified in
   the ESO SPY survey. We performed NLTE model atmosphere analyses and found
   that the PG1159 star is a low-gravity, extremely hot (\Teff=160\,000\,K,
   \logg=6) star, having a carbon-helium dominated atmosphere with considerable
   amounts of oxygen and neon (He=38\%, C=54\%, O=6\%, Ne=2\% by mass). It is
   located within the planetary nebula nuclei instability strip, hence, future
   searches for an associated PN as well as for stellar pulsations might be
   successful. The DO white dwarf is a unique object. From the relative strength
   of neutral and ionized helium lines we found \Teff$\approx$60\,000\,K,
   however, the \ion{He}{ii} lines are extraordinarily strong and cannot be
   fitted by any model. Similar problems were encountered with hot subdwarfs and
   white dwarfs showing signatures of a super-hot wind. The reason is unknown.
        }
   \maketitle

\begin{figure*}[tbp]
\includegraphics[width=\hsize]{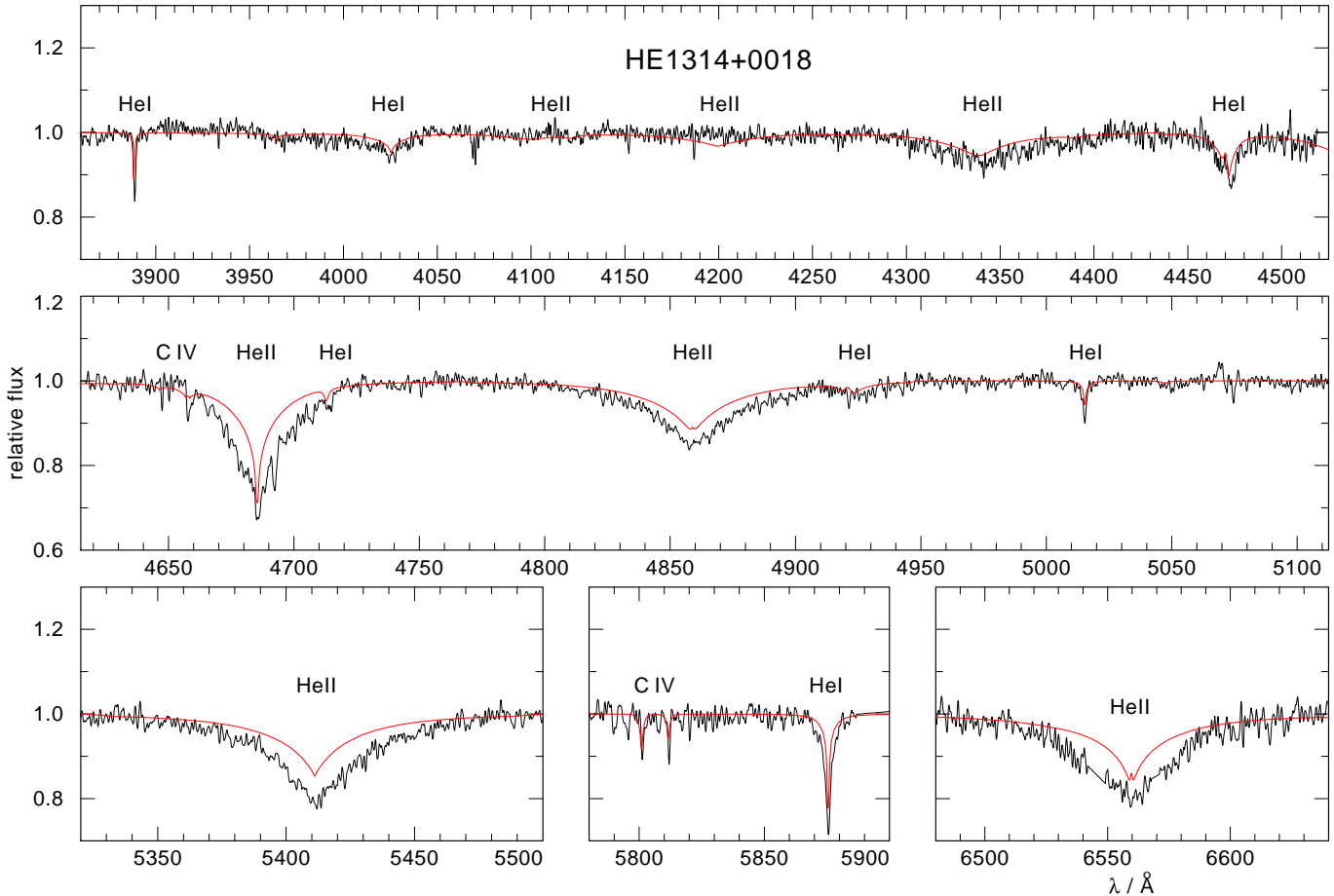}
  \caption[]{
Spectrum of the new DO white dwarf. Overplotted is a model with \Teff=60\,000\,K,
 \logg=7.5, and C/He=0.001. The \ion{He}{i} lines fit well, but all the
 \ion{He}{ii} lines in the model are much too weak. No acceptable fit to the
 complete spectrum is found. The synthetic spectrum was degraded to the instrumental resolution.
 }
  \label{he1314_overview}
\end{figure*}

\begin{figure*}[tbp]
\includegraphics[width=\hsize]{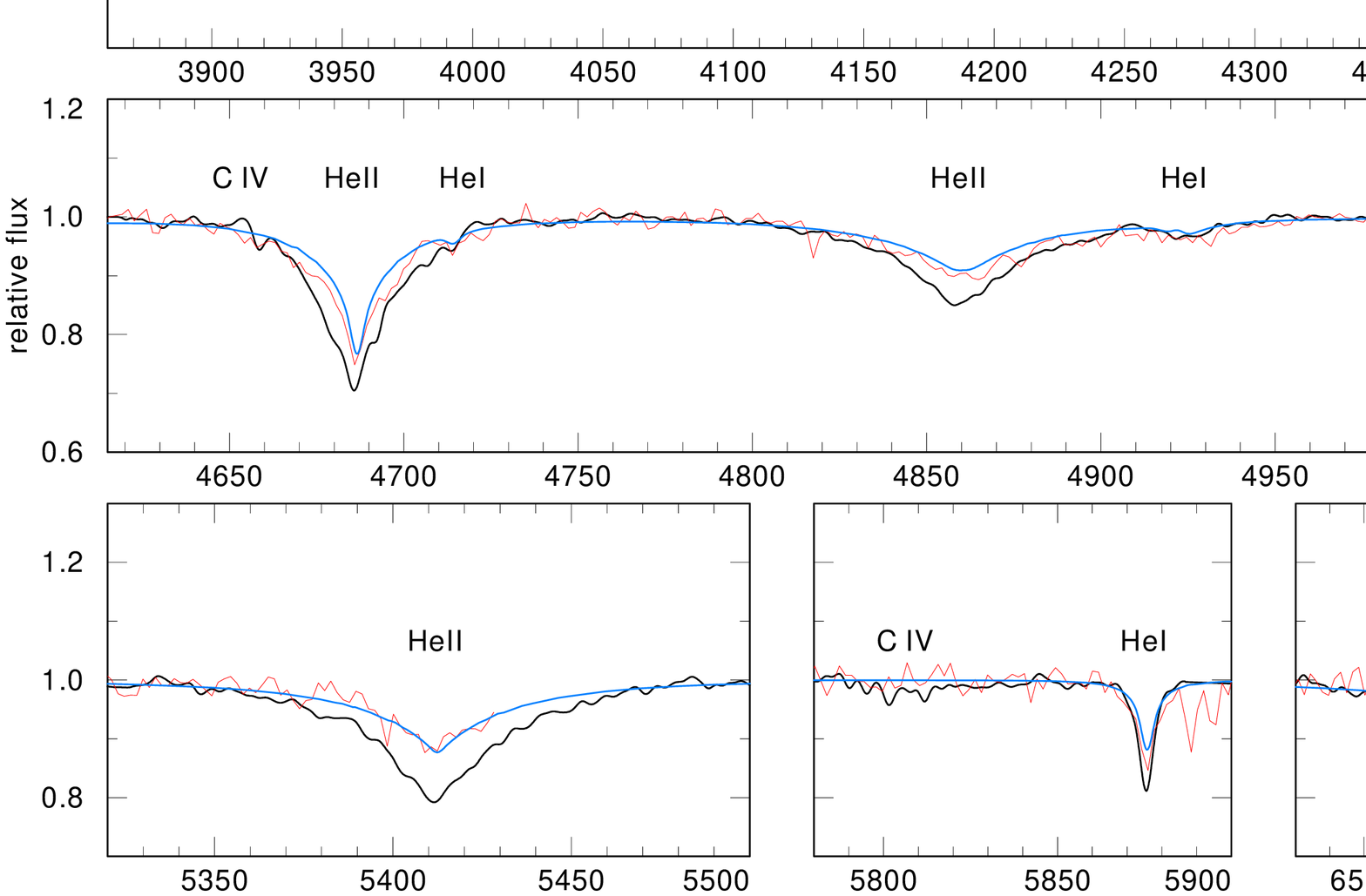}
  \caption[]{
\dostar\ (thick line) compared to Lanning 14 (thin line), which is a DO with
similar parameters. The smooth thin line
is a model atmosphere with \Teff=60\,000\,K and \logg=8 and fits well to
Lanning~14 (\Teff=58\,000\,K, \logg=7.9). The \ion{He}{ii} lines in \dostar\ are
significantly deeper and cannot be fitted with any model spectrum. The \dostar\
spectrum and the model spectrum have been degraded according to the resolution
of the Lanning~14 spectrum (3\,\AA).
 }
  \label{he1314_compared_lanning14}
\end{figure*}

\begin{figure*}[tbp]
\includegraphics[width=\hsize]{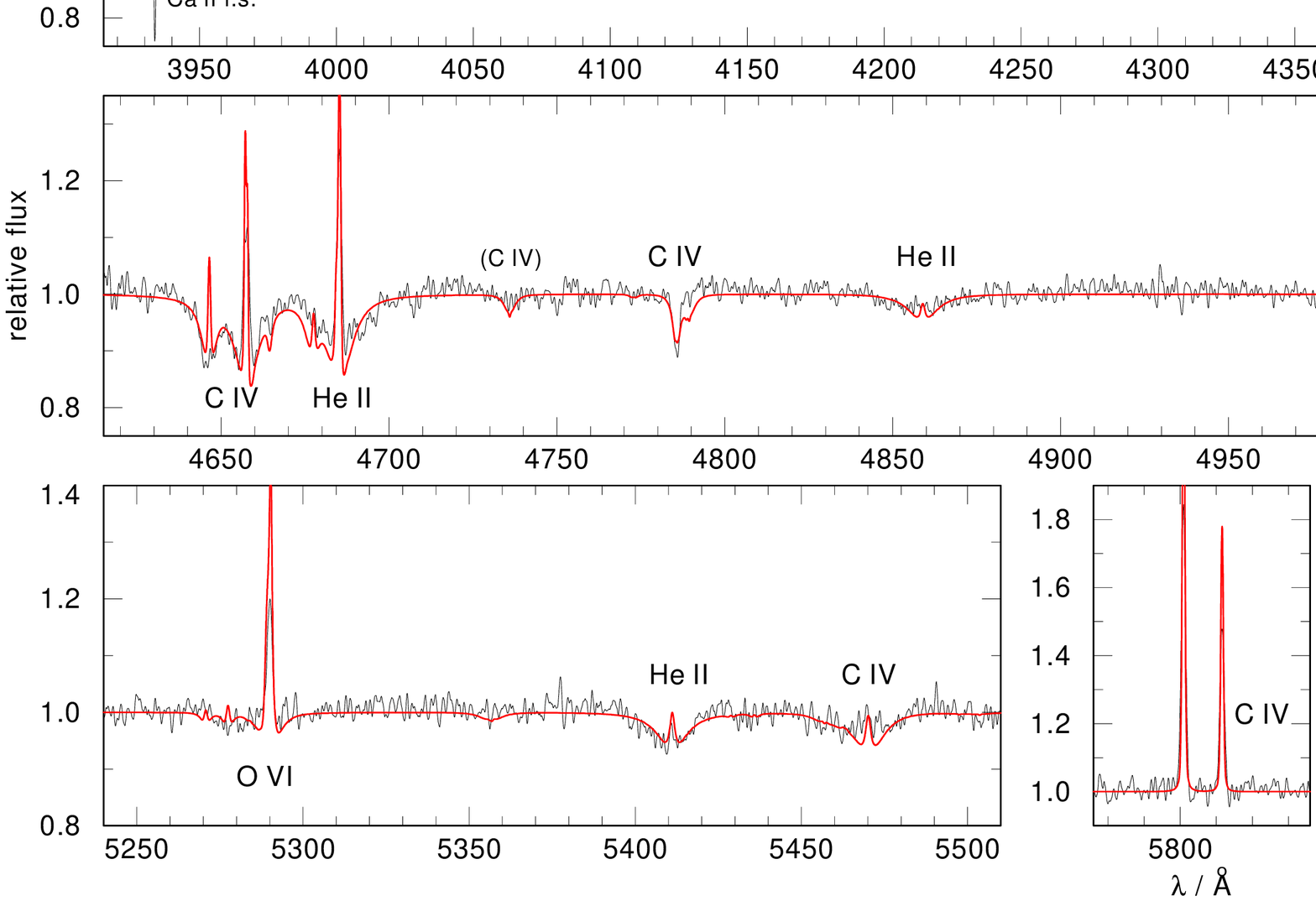}
  \caption[]{
Spectrum of the new PG1159 star with the best fit model overplotted. An
 expanded view of the 3627--3662\,\AA\ and 3808-3914\,\AA\ regions comprising neon
 lines are shown in Figs.\,\ref{he1429_ne7_singlet} and
 \ref{he1429_ne7_triplet}. ``X'' denotes an artifact. There is an unidentified
absorption line at 4295\,\AA.The observed spectrum was smoothed with a
 Savitzky-Golay filter (Press \etal 1992) and the synthetic spectrum was degraded to the instrumental resolution.
 }
  \label{he1429_overview}
\end{figure*}

\section{Introduction} 

   The ESO Supernovae Ia Progenitor Survey (SPY) is aimed at
   finding binary white dwarfs to test the double-degenerate (DD) scenario for supernova type Ia progenitors
   (Napiwotzki \etal 2001, 2003). Multi-epoch high-resolution
   spectroscopy of almost 1000 white dwarfs has been obtained
   with the UV-Visual Echelle Spectrograph (UVES; Dekker et al.
   2000) mounted at the UT2 telescope (Kueyen) of the ESO VLT,
   allowing identification of DDs by radial velocity variations.

   Other interesting objects found in the SPY project are, e.g.,
   subdwarf B stars (Lisker \etal 2004), and DB
   white dwarfs (Christlieb \etal 2004, in prep.). Here we
   report on the identification and spectrum analysis of a PG1159 star and a DO white
   dwarf.

PG1159 stars are hot hydrogen-deficient (pre-) white dwarfs ($T_{\rm
eff}$=75\,000--200\,000\,K, $\log g$=5.5--8 [cm/s$^2$]; Werner 2001). They
are perhaps the outcome of a late helium-shell flash, a phenomenon that drives
the currently observed fast evolutionary rates of three well-known objects
(FG~Sge, Sakurai's object, V605 Aql). Flash-induced envelope mixing produces a
H-deficient stellar surface. The photospheric composition then essentially
reflects that of the region between the H- and He-burning shells in the
precursor AGB star. The He-shell flash transforms the star back to an AGB star
and the subsequent, second post-AGB evolution explains the existence of
Wolf-Rayet central stars of planetary nebulae and their successors, the PG1159
stars.

DO white dwarfs are hot helium-rich white dwarfs (\Teff=47\,500--120\,000\,K,
\logg=7--8; Dreizler \& Werner 1996), which are thought to be the progeny of
PG1159 stars. After entering the white dwarf (WD) cooling sequence, heavy
elements are removed from the PG1159 atmospheres by gravitational settling when
the luminosity decreases during their evolution, leaving behind the lightest
element, helium, and transforming the star into a DO white dwarf.

Although this evolutionary scenario seems to be well established, many unsolved
problems remain, which justifies a closer look at new PG1159 stars
and DO white dwarfs. They are relatively rare objects when compared to the total
number of white dwarfs and central stars known, often with quite remarkable
individual characteristics.

We present here a newly discovered He-rich white dwarf with unusually strong
spectral lines. We also present a new PG1159 star which belongs to the rare
subgroup of highly luminous objects. In Section~2 we describe the observations
and then turn to the spectral analysis of the two stars in Section~3. We
conclude with a discussion in Section~4.

\section{Observations} 

\dostar\ ($\alpha_{2000}=13^{\rm h}17^{\rm m}24\fs 74$, $\delta_{2000}=+00\degr
02' 37\farcs 1$, B=15.6\,mag) and \pgstar\ ($\alpha_{2000}=14^{\rm h}32^{\rm
m}20\fs 69$, $\delta_{2000}=-12\degr 22' 47\farcs 9$, B=15.8\,mag)  were
identified in the Hamburg ESO survey (HES; Wisotzki \etal 2000, Christlieb \etal
2001) as white dwarf candidates and were therefore included in the SPY
project. \dostar\ (=PG1314+003) was classified as ``sdOC'' in the Palomar-Green
catalog (Green \etal 1986). Based on the SPY spectra presented here, \dostar\ is listed in
Christlieb \etal (2001) as DO white dwarf. Later,
it was also identified in the Sloan Digital Sky Survey (Krzesi\'nski \etal 2004).
\pgstar, independently re-discovered by us, was classified
by Kilkenny \etal (1997) as ``cont'' or ``sdO?'' (see their
Table~1). The EC survey identifier of this object is
EC\,14296-1209. The high-resolution UVES spectra  obtained in the
course of the SPY project revealed that \pgstar\ is actually
a PG1159 star.
 
The spectra presented here were taken between April 2000 and July 2002. The
integration time for each spectrum was 10 minutes. \dostar\ was observed three
times and \pgstar\ twice. The respective spectra were co-added to improve the
S/N ratio.  The SPY instrument setup (Dichroic~1, central wavelengths 3900\,\AA\
and 5640\,\AA) used UVES in a dichroic mode and nearly complete spectral
coverage from 3200\,\AA\ to 6650\,\AA\ with only two $\approx$80\,\AA\ wide gaps
at 4580\,\AA\ and 5640\,\AA\ is achieved. SPY was implemented as a service mode
program, which took advantage of those observing conditions, which are not
usable by most other programs (moon, poor seeing, clouds).  A wide slit
($2\farcs 1$) is used to minimize slit losses and a $2\times 2$ binning is
applied to the CCDs to reduce read-out noise. The wide slit reduces the spectral
resolution to $R=18\,500$ (0.2\,\AA\ at 3600\,\AA) or better, if seeing disks
were smaller than the slit width.

The spectra were reduced with a procedure developed by Karl (2004, in prep.)
using the ESO MIDAS software package, partly based on routines from the UVES
pipeline reduction software provided by ESO (Ballester et al.\ 2003). Since the
sampling of the spectra is much higher than needed for our purpose, we rebinned
them to 0.1\,\AA\ stepsize. This produces only a slight degradation of the
resolution while considerably improving the signal-to-noise level.
Figures~\ref{he1314_overview} and \ref{he1429_overview} display the spectra of
the DO white dwarf and the PG1159 star, respectively.

\begin{figure}[tbp]
\includegraphics[width=\columnwidth]{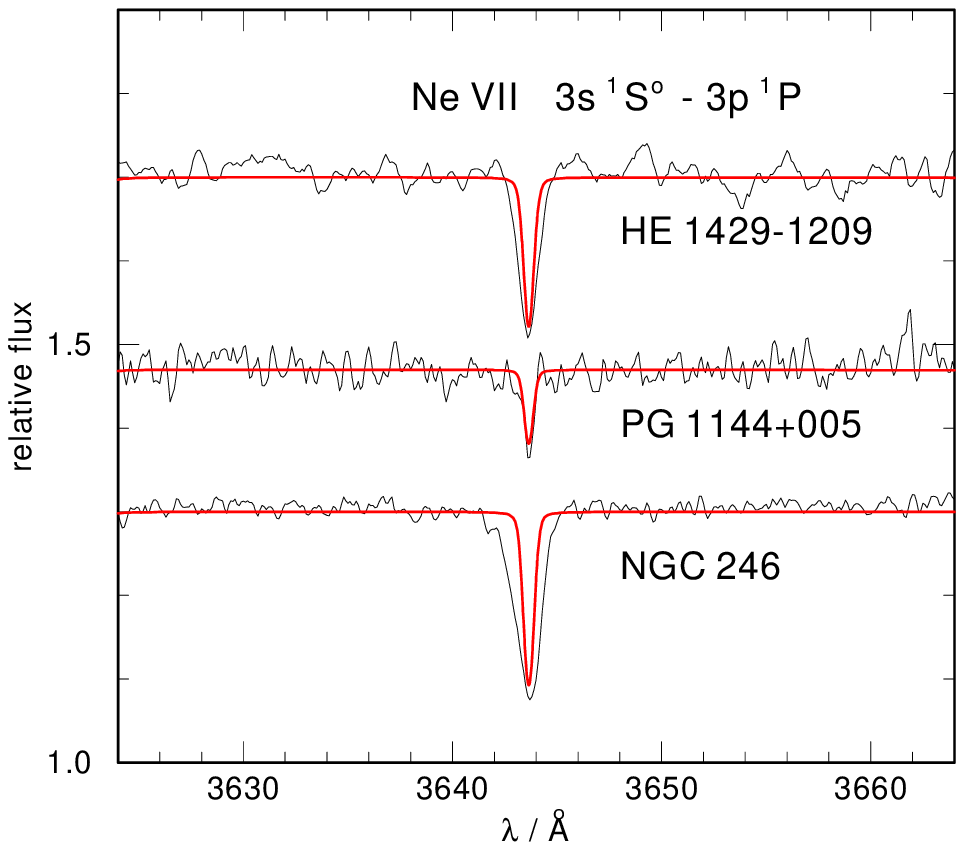}
  \caption[]{
Detection of the \ion{Ne}{vii}~3644\,\AA\ singlet in \pgstar. Overplotted is our
 best fit model with a Ne abundance set to 2\% by mass. For comparison, we show spectra of two other PG1159 stars
 together with their models (assuming also Ne=2\%; presented in Werner \etal 2004b).
 }
  \label{he1429_ne7_singlet}
\end{figure}

\section{Spectral analysis} 

Line blanketed non-LTE model atmospheres were computed using our {\sc PRO2} code
(Werner \etal 2003). The models assume plane-parallel geometry and hydrostatic
and radiative equilibrium.

The atomic models employed for the calculation of PG1159 stellar atmosphere
models account for the most abundant elements in PG1159 stars, namely,  helium,
carbon, oxygen, and neon. For more details we refer to Werner \etal (2004a,
2004b). The atomic models for DO atmospheres comprise very detailed \ion{He}{i}
and \ion{He}{ii} ions. Essentially, the input physics of the models is similar
to that described in Dreizler \& Werner (1996). The model atom for the
calculation of carbon line profiles in the DO is similar to that used for the
PG1159 star and is consistently included in the model structure calculations.

\subsection{The DO star \dostar}

The spectrum of \dostar\ exhibits lines from neutral and ionized helium, i.e.\
we have detected a ``cool'' DO (``hot'' DOs show only \ion{He}{ii} lines). This
usually allows an accurate determination of \Teff\ and \logg\ (Dreizler \&
Werner 1996). We also detect the \ion{C}{iv}~5801/5812\,\AA\ doublet and a very
weak feature of this ion close to \ion{He}{ii}~4686\,\AA.  A first guess for the
photospheric parameters can be obtained by comparison with a compilation of many
other DO spectra in the reference just mentioned. Based on this, we computed a
small grid of pure helium models with \Teff=55\,000--80\,000\,K and
\logg=7--9. The \ion{C}{iv} lines are indicative for a C abundance of the order
0.1\% by number.

Surprisingly, we are unable to obtain an acceptable fit to the observed
spectrum. Figure~\ref{he1314_overview} shows the best fit model to the
\ion{He}{i} lines, which is obtained with \Teff=60\,000\,K and \logg=7.5. It is
obvious that the synthetic \ion{He}{ii} lines are much too weak compared to the
observation. We found no model that can fit the strong \ion{He}{ii} lines.  If
we increase \Teff, these lines become slightly deeper, but never reach the
observed equivalent widths. \Teff$>$70\,000\,K can be excluded because the
\ion{He}{i} lines disappear. \Teff$<$55\,000\,K is excluded because the
\ion{He}{i} lines become too strong and the \ion{He}{ii} lines are getting even
weaker. A compromise is reached at about 65\,000\,K in a sense that the relative
\ion{He}{i}/\ion{He}{ii} line strengths in the model are similar to the observed
ones. But, clearly, lines from both ionization stages are much too weak. This
becomes also evident if we compare the spectrum of \dostar\ directly to that of
other DOs in this temperature range (e.g.\ in Dreizler \& Werner 1996). In
Fig.\,\ref{he1314_compared_lanning14} a comparison to Lanning~14 is shown. This
DO has parameters very similar to our new DO.

As already indicated, this situation was never encountered before in analyses of
cool DOs. However, this problem is reminiscent of what we faced with those
``hot'' DOs that show signatures of a super-hot wind.  We have found that a
large fraction (50\%) of the hot DO white dwarfs shows such signatures in the
optical spectrum (Werner \etal 1995, Dreizler \etal 1995a). These hot stars also
show high ionization absorption lines of the CNO elements (e.g.\ \ion{C}{vi},
\ion{N}{vii}, \ion{O}{viii}, and even \ion{Ne}{x}). The high excitation
potentials involved require temperatures approaching almost 10$^6$\,K and the
triangular shaped line profiles suggest their formation in a rapidly
accelerating wind from the WD, reaching a terminal velocity of the order of
10\,000\,km\,s$^{-1}$. At the same time, all \ion{He}{ii} lines have symmetric
profiles and, like in the present case, they are much too strong to be fitted by
any DO model. The reason for this simultaneous occurrence of ``hot-wind''
signatures and too-strong \ion{He}{ii} lines in hot DOs is unknown.

Our new DO star also has too strong \ion{He}{ii} lines but, in contrast, shows
no signatures of super-high ionization metal lines. But on the other hand these
signatures are rather weak in some of the ``hot-wind'' DOs. So we can speculate
that in the present case the physical reason for the excessive line strengths is
the same like in the ``hot-wind'' DOs, but the respective super-high ionization
metal lines remain undetected here. The low effective temperature of \dostar\ in
comparison to the ``hot-wind'' DOs might be linked to this non-detection.

We stress that this wind phenomenon along with too-strong lines is not
restricted to DO white dwarfs. We have also detected one DA white dwarf with
similar high-ionization features (Dreizler \etal 1995a) and much too deep
hydrogen Balmer lines when compared to models. This is the only case discovered
among DA white dwarfs so far, and the question arises why there is such a strong
preference among DOs to show this phenomenon.

A similar problem with
inconsistent line-profile fits was encountered in analyses of some sdB and sdOB
stars, in the range \Teff=40\,000\,--60\,000\,K and \logg$\approx$6 (Heber \etal
1999). The temperature derived from the \ion{He}{i}/\ion{He}{ii} ionization
balance does not fit the strength of the hydrogen Balmer lines. The Balmer
lines are too strong, in analogy to our case presented here, where we have
too strong \ion{He}{ii} lines. The Balmer lines in these subdwarfs can be fitted
by models with \Teff\ {\it lower} than derived from the \ion{He}{i}/\ion{He}{ii}
ionization balance, whereas in contrast the \ion{He}{ii} lines in our DO white
dwarf in tendency require a {\it higher} \Teff. Whether the solution to these
problems with inconsistent line-profile fits in white dwarfs and subdwarfs can be
explained by strong line blanketing (e.g.\ by iron-group elements) remains to be
investigated.

We have also considered the possibility that \dostar\ shows a binary composite
spectrum. This can be ruled out because, as already mentioned, the strong
\ion{He}{ii} lines cannot be matched by any DO model, even if we disregard the
presence of \ion{He}{i} lines.

From the \ion{C}{iv}~5801/5812\,\AA\ doublet we find a carbon abundance of
C/He=0.003 by number, under the assumption of \Teff=60\,000\,K and \logg=8.

\subsection{The PG1159 star \pgstar}

A comparison of the \pgstar\ spectrum with other PG1159 stars immediately
reveals that we have found another very hot, low-gravity PG1159 star which
is rather similar to the well known central star of the planetary nebula
NGC\,246 and the central star \rxj\ (Rauch \& Werner 1997). The latter has
\Teff=170\,000\,K and \logg=6.0 (Werner \etal 1996). According to the
classification scheme introduced by Werner (1992), the spectral subtype is
``lgE'' (meaning low-gravity star with emission lines). The spectrum is
characterized by lines from highly ionized species (\ion{He}{ii}, \ion{C}{iv},
\ion{O}{vi}, Fig.\,\ref{he1429_overview}), often present as absorption lines
with central emission reversals or even as pure emission lines. The absorption
features in ``lgE'' subtypes are narrower than in other PG1159 stars pointing at
a lower surface gravity, hence, higher stellar luminosity.

We have calculated a small grid of model atmospheres in the range 140--180\,kK
and \logg=5.7--6.5, assuming an element composition of He=38\%, C=54\%, O=6\%,
Ne=2\% (by mass). This is the composition found for \rxj. Having found the best
fit model for \pgstar, we vary the element abundances, guided by  comparison of
selected lines in the observed and synthetic spectra. No attempt is made to
perform a more formal fit, because that would require an extensive model grid
varying six parameters. Our error estimates are conservative in order to account
for this approach.

\begin{figure*}[tbp]
\includegraphics[width=\textwidth]{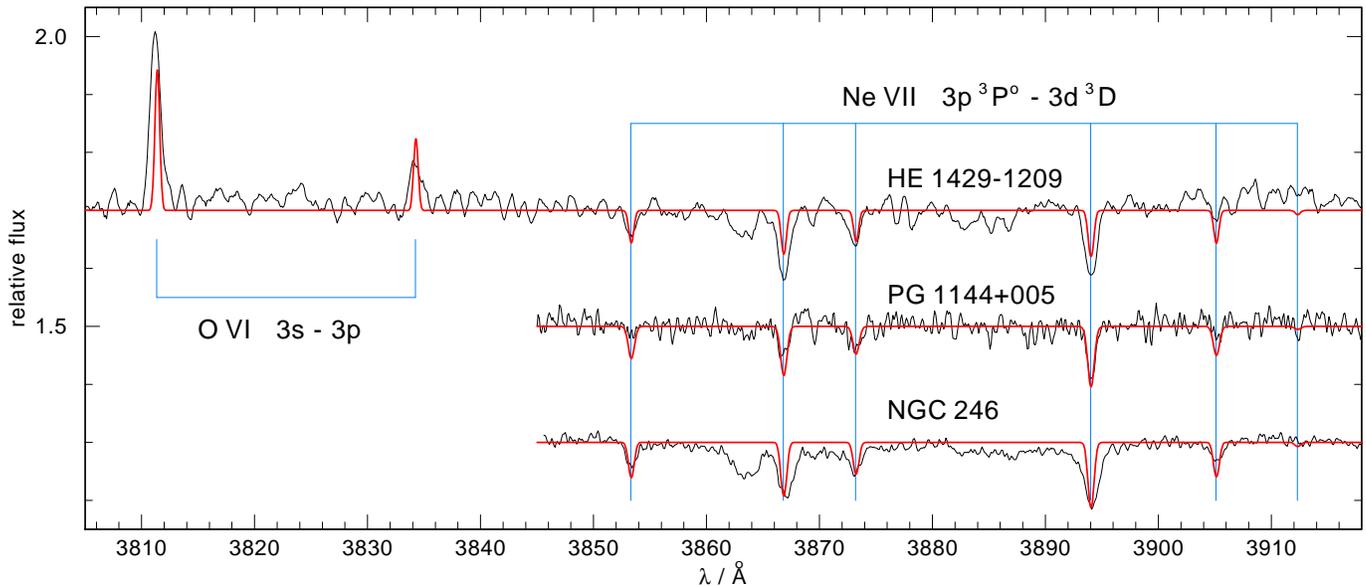}
  \caption[]{
Detection of a \ion{Ne}{vii} multiplet in \pgstar. For
comparison, 
we show spectra of two other PG1159 stars. Synthetic spectra are from the same
models as in Fig.\,\ref{he1429_ne7_singlet}. There is an unidentified
absorption line at 3862\,\AA.
 }
  \label{he1429_ne7_triplet}
\end{figure*}

We stress in particular the presence of a \ion{Ne}{vii} line at 3644\,\AA\
(Fig.\,\ref{he1429_ne7_singlet}).  We have recently performed a systematic
investigation about neon in a large number of PG1159 stars (Werner \etal 2004b)
and found that such a prominent \ion{Ne}{vii} line is preferentially exhibited
by ``lgE'' subtypes. It has been shown that neon is strongly overabundant (about
20 times solar, i.e., 2\% by mass), otherwise the line would be
undetectable. The strength of this line is similar to that in \ngc, confirming
that the neon abundance in \pgstar\ is also of the order 2\%. In the paper just
cited, we have identified for the first time a \ion{Ne}{vii} multiplet in the 3850--3910\,\AA\
range. Figure~\ref{he1429_ne7_triplet} shows that this multiplet is also present
in our new PG1159 star. Generally, the computed equivalent widths of the
\ion{Ne}{vii} lines in \pgstar\ appear underestimated, as it is the case in
\ngc, suggesting that the neon abundance may be even higher than 2\%.

Our best fitting model essentially confirms our first guess for the atmospheric parameters. We
find:
\begin{eqnarray*}
\Teff   &=& 160\,000\,{\rm K} \pm 15\,000\,{\rm K} \qquad \logg=6.0 \pm 0.3 {\rm  \ \ [cm/s}^2{\rm ]}\\
{\rm He}&=&38\% \quad  
{\rm C} =54\% \quad   
{\rm O} =6\% \quad  
{\rm Ne}=2\%
\end{eqnarray*}
The estimated error for abundances is 0.3\,dex.
Stellar mass and luminosity can be derived by comparing the star's position in
the \logg--$\log$\,\Teff\ diagram with theoretical evolutionary tracks. We use the post-AGB
He-burner tracks of Wood \& Faulkner (1986; Fig.\,\ref{gteff}) and derive: 
\begin{eqnarray*}
M/{\rm M}_\odot&=&0.68^{+0.15}_{-0.08} \qquad \log L/{\rm L}_\odot=4.04^{+0.09}_{-0.05}\\
d/{\rm kpc}&=&5.2^{+1.6}_{-2.2}
\end{eqnarray*}
\pgstar\ is still very luminous, i.e., on the horizontal part of the post-AGB
evolutionary track in the HRD, located within the domain of the hottest central
stars of planetary nebulae. The spectroscopic distance was obtained by comparing
the visual flux (V=16.1\,mag) with the flux of the final model:
H$_\nu$[5400\,\AA]=$2.85 \cdot 10^{-3}$ erg/cm$^2$/s/Hz). Interstellar reddening
is probably small and was neglected for this determination.

\section{Conclusions}

The new He-rich white dwarf joins the group of 33 DO white dwarfs (see
Dreizler \& Werner 1996 for an analysis of most of them). It belongs to the
subgroup of ``cool'' DO stars which exhibit both ionized and neutral helium
lines. In contrast to all other cool DOs, \dostar\ shows extraordinarily strong
\ion{He}{ii} lines, which cannot be matched by any model atmosphere. This
phenomenon was only known from the ``hot'' DOs (i.e. those exhibiting
exclusively \ion{He}{ii} lines). However, in these cases it simultaneously
occurs along with  ultra-high ionization metal lines, probably stemming from a
super-hot wind. No such signatures are seen in \dostar. The extreme
strength of the \ion{He}{ii} lines is not understood.

The effective temperature of \dostar\ is close to 60\,000\,K. We found
C/He=0.003 by number. This means that \dostar\ is the coolest DO with a safe
detection of a photospheric trace metal. The occurrence of metals in DOs is
thought to be radiatively supported against gravitational settling. \dostar\ is
obviously located at the cool end of the \Teff-range, beyond which radiation
forces can no longer keep carbon up in the atmosphere.  This provides a
constraint for detailed self-consistent  NLTE diffusion models for DO white
dwarfs, which still need to be computed.

\begin{figure}[tbp]
\includegraphics[width=\columnwidth]{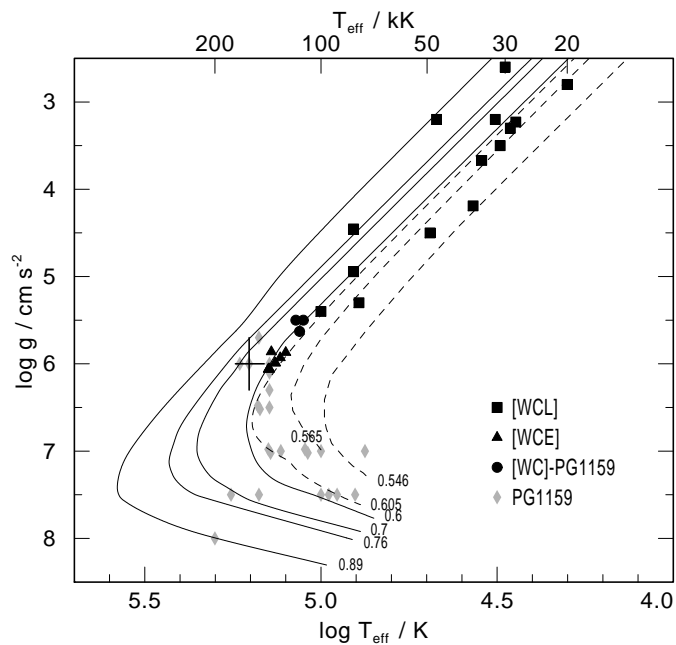}
  \caption[]{
Position of the new PG1159 star \pgstar\ (with error bars) in the \logg--$\log$\,\Teff\ diagram. We
identify Wolf-Rayet central stars (Hamann 1997), PG1159 stars as well as
[WC]--PG1159 transition objects. Evolutionary tracks are from Sch\"onberner
(1983) and Bl\"ocker (1995) (dashed lines), and Wood \& Faulkner (1986) (labels:
mass in M$_\odot$). For details on other objects than \pgstar\ see Werner (2001).
 }
  \label{gteff}
\end{figure}

The new PG1159 star is among the hottest objects within this group,
which now comprises 32 stars (30 of them are listed in Dreizler \etal 1995b). It
is very similar to six members (having subtype lgE), which are low-gravity,
i.e., high-luminosity objects, being still on the hot, horizontal part of their
post-AGB evolutionary track in the HRD. It is thus not surprising that all
except one of these previously known lgE-PG1159 stars have an associated
planetary nebula, which are rather old and large, but not yet dispersed as
is the case for most of the more evolved PG1159 stars (having entered the white
dwarf cooling sequence).  So we expect that our new PG1159 star could also be
surrounded by an extended PN.  This prompted us to perform a search for faint
nebulosities on the Palomar Observatory Sky Survey (POSS) plates and the Southern H-Alpha
Sky Survey Atlas ({\sc SHASSA}; Gaustad \etal 2001).  We could not find anything
resembling a PN on the POSS plates, but an elongated faint nebulosity, extending
over several degrees, is visible in the {\sc SHASSA} image. However, the
structure is not centered on \pgstar. It is possible that this is the \pgstar\
PN in a very advanced stage of interaction with the interstellar medium, but
deeper images would be necessary to confirm this.  In the \logg--\Teff\ diagram
(Fig.\,\ref{gteff}) \pgstar\ is located close to the PNNVs (PN nuclei variables)
\keins, \rxj, NGC\,246, and Longmore~4. It could well be, that a photometric
study will show that the new PG1159 star is also a pulsator. Concerning the
atmospheric abundances, \pgstar\ is a ``typical'' PG1159 star. It further
confirms that these objects present intershell matter of the progenitor AGB star
on their surface, possibly as a result of a late He-shell flash.

\begin{acknowledgements}
We thank Ronald Weinberger (Innsbruck) for helpful discussions. T.R.\ is
supported by the DLR under grant 50\,OR\,0201. R.N.\ acknowledges support by a
PPARC advanced fellowship. C.K.\ gratefully acknowledges financial
support by the Deutsche Forschungsgemeinschaft through grant 
NA365/2-2. We are grateful to the ESO staff
at Paranal for carrying out the VLT/UVES  Service Mode observations. This
article made use of the Southern H-Alpha Sky Survey Atlas ({\sc SHASSA}), which
is supported by the National Science Foundation.
\end{acknowledgements}

\end{document}